\documentclass[useAMS]{mn2e}
\usepackage{times}
\input{epsf}

\def\lhs{$\ell_{\rm h}/\ell_{\rm s}$}
\def\lnth{$\ell_{\rm nth}/\ell_{\rm h}$}

\title[]{Patterns of energy-dependent variability from Comptonization}

\author[M. Gierli\'nski and A. A. Zdziarski]
{Marek~Gierli\'nski$^{1,2}
$\thanks{E-mail:Marek.Gierlinski@durham.ac.uk}
and Andrzej A. Zdziarski$^{3}$\\
$^1$Department of Physics, University of Durham, South Road,
Durham DH1 3LE, UK\\
$^2$Astronomical Observatory, Jagiellonian University, Orla 171,
30-244 Krak{\'o}w, Poland\\
$^3$Centrum Astronomiczne im.\ M. Kopernika, Bartycka 18, 00-716 Warszawa, Poland
}

\date{Submitted to MNRAS}
\pagerange{\pageref{firstpage}--\pageref{lastpage}} \pubyear{2005}

\begin{document}

\topmargin = -0.5cm

\maketitle

\label{firstpage}

\begin{abstract}
We study fractional variability as a function of energy from black-hole
X-ray binaries on timescales from milliseconds to hundreds of seconds.
We build a theoretical model of energy-dependent variability in which
the X-ray energy spectrum varies in response to a changing physical
parameter. We compare these models to rms spectra obtained from {\it
RXTE\/} PCA observations of black-hole binaries XTE~J1550--564 and
XTE~J1650--500. We show that two main variability models are consistent
with the data: variable seed photon input in the hard state and
variable power in the Comptonized component in the soft and very high
states. The lack of clear reflection features in the rms spectra
implies that the reflection and the X-ray continuum, when integrated
over Fourier frequencies, are correlated and vary with similar
fractional amplitudes. Our models predict two important features of rms
spectra, not possible to be clearly seen by the PCA due sensitivity
limits. At soft X-rays, $\la$3 keV, we predict the presence of a break
in the rms spectrum at energy directly related to the seed photon
temperature. At higher energies, $\sim$20--30 keV, we predict a peak in
the rms spectrum originating from the variability of the spectrum
produced by a hybrid thermal/non-thermal electron distribution. If
these features are confirmed by broad-band observations, they will
impose important constraints on the origin of the seed photons for
Comptonization and the electron distribution in the hot plasma.
\end{abstract}


\section{Introduction}
\label{sec:introduction}

X-ray emission from accreting black holes is commonly thought to
originate from inverse Compton scattering of cooler disc photons in a
hot optically thin plasma. Depending on the geometry of the accretion
flow and the distribution of power between the disc and the hot plasma,
a variety of spectral distributions can be produced. This translates
into a variety of observed spectral states. One approach to
understanding the physics of accretion is by fitting various models to the
time-averaged energy spectra. The spectral decomposition of the data is
fairly well understood, and typically requires a model consisting of
disc emission, its Comptonization and Compton reflection of the
hard X-ray photons from the disc (see e.g. Zdziarski \& Gierli{\'n}ski
2004 and references therein).

Another approach to the X-ray data is by analysing their variability on
various timescales. Fast aperiodic variability on the timescales from
milliseconds to hundreds of seconds is often studied using power
density spectra (PDS), in particular, by tracing quasi-periodic
oscillations (QPO). For a review, see van der Klis (2004). Most of this
variability occurs on dynamical timescales in the inner part of the
accretion flow, so it makes an excellent probe of the deep
gravitational potential around the compact object. However, despite
huge amounts of available data, we are still far from understanding of
{\em how} the rapid X-ray variability is produced. Many of existing
models propose oscillations in the accretion disc (e.g. Cui, Zhang \&
Chen 1998; Titarchuk, Osherovich \& Kuznetsov 1999; Psaltis \& Norman
2000) as the origin of variability. Though these models have been
successful in explaining observed characteristic frequencies, it is not
clear how oscillations of the disc are translated into varying X-rays.
Giannios \& Spruit (2004) proposed that QPOs in the inner hot flow can
be excited by interaction between the flow and the outer cold disc
(e.g.\ by cooling-heating feedback loop). Bursa et al.\ (2004)
suggested that hard X-rays can be modulated via gravitational lensing
of the oscillating accretion flow in the vicinity of the black hole.
Another suggestion involves dense cold blobs of material drifting
through the inhomogeneous hot inner flow and providing with a variable
source of the seed photons for Comptonization (B{\"o}ttcher \& Liang
1999). An alternative set of models that can directly explain the
origin of X-ray modulation invokes propagation of X-ray flares in the
accretion flow (e.g. B{\"o}ttcher \& Liang 1998; Poutanen \& Fabian
1999; {\.Z}ycki 2003).

There are a few possibilities of bridging over the fairly well
understood energy spectra and the enigmatic variability. One of them is
to study how X-ray energy spectra change with the Fourier frequency, by
means of the so-called frequency-resolved spectroscopy. A handful of
bright objects have been studied this way, both containing black holes
(Revnivtsev, Gilfanov \& Churazov 1999b, 2001) and neutron stars
(Gilfanov, Revnivtsev \& Molkov 2003). A number of important
conclusions have been obtained from these data. One particularly
interesting result is that the strength of Compton reflection in the
low/hard state of black-hole binaries significantly decreases with the
increasing Fourier frequency.

Another promising but not yet very well studied field of research is
the energy dependence of rapid X-ray variability. It can bridge over
the fairly well understood energy spectra and the enigmatic
variability. The simplest approach is by looking at the fractional root
mean square variability amplitude (integrated over a range of
frequencies or timescales) as a function of energy, rms($E$), or in
other words, a relative variability spectrum. Such a spectrum can tell
us about how the spectral components (disc or Comptonization) vary with
respect to each other and whether they change their spectral shape on
the observed timescales. Variability spectra have been recently
obtained both from Galactic and supermassive black holes (e.g.
Revnivtsev, Borozdin \& Emelyanov 1999a; Lin et al.\ 2000;
Wardzi{\'n}ski et al.\ 2002; Vaughan \& Fabian 2004). Also, the
amplitude of QPOs as a function of energy has also been studied (e.g.
Rao et al.\ 2000; Gilfanov et al.\ 2003; Rodriguez et al.\ 2004a, b).
However, until recently, little theoretical interpretation was given.
Zdziarski (2005) proposed a theoretical model considering radial
dependence of the local variability in the disc. By assuming the local
rms and disc temperature decreasing with increasing radius, he found an
rms($E$) increasing with energy, consistent, e.g., with an ultrasoft
state of GRS 1915+105. Zdziarski et al.\ (2002), hereafter Z02,
analysed the spectral variability of Cyg X-1 from the {\it RXTE}/ASM and
{\it CGRO}/BATSE on timescales of days and months, and reported two
distinct patterns of rms($E$) in different spectral states. They
proposed that in the hard state the variability was driven by changes
in the inner radius of the truncated disc, which in turn varied the
seed photon input for Comptonization. In the soft state, it was brought
about by variations of the power released in a hot corona above the
disc.

In this paper, we investigate patterns of rms($E$) variability
generated by variations of physical properties of the accretion flow.
We consider a particular spectral model of a hybrid,
thermal/non-thermal Comptonization and study the effects of varying
parameters of this model. We compare the results with rms spectra of
rapid X-ray variability of two Galactic black holes, XTE J1650--500 and
XTE J1550--564.

We would like to point out that our variability spectra have their own
limitations. First, we have chosen to integrate them over a wide range
of Fourier frequencies (those available to the {\it RXTE}/PCA
instrument, see Section \ref{sec:data} below), while the spectral
dependence on frequency has been shown to be important in some cases,
see, e.g., Revnivtsev et al.\ (1999b). We note that it is entirely
possible to create rms($E$) integrated over a narrow range of
frequencies, and then that rms($E$) times the average spectrum will be
equal to the corresponding frequency-resolved spectrum. Such an
analysis is, however, beyond the scope of the present work because it
would increase the dimension of the parameter space, thus leading to a
substantial increase of the complexity of the study.  Also, that
approach requires data of significantly higher statistics than those
required in the case of integration over all available frequencies.
Consequently, the results on interpretation of X-ray data presented
here are mostly valid for the range of frequencies dominating the power
spectrum (typically $\sim$0.1--10 Hz, see Section \ref{sec:patterns}
below). On the other hand, our theoretical results are general, and can
be directly applied to frequency-resolved rms($E$) in future work.

Furthermore, neither rms($E$) nor frequency-resolved spectra carry
information about either phase/time lags (for a review see Poutanen
2001 and references therein) or coherence (Vaughan \& Nowak 1997)
between signals at different energies at a given Fourier frequency.
Also, our theoretical interpretation of rms($E$) is based on a
one-component model of Comptonization. Thus, this approach does not
deal with, e.g., propagation effects, likely to be important in actual
flows (e.g. Kotov, Churazov \& Gilfanov 2001).

\section{Data reduction}
\label{sec:data}

We have analysed several observations (with unique identifiers, called
obsids) of XTE J1650--500 and XTE J1550--564 from the {\it RXTE\/} PCA and
HEXTE detectors. For data reduction, we used {\sc ftools} 5.3. We
extracted energy spectra from the top layer of the detector 2 of the PCA and
added a 1 per cent systematic error in each channel. We extracted HEXTE
spectra from both clusters. For energy spectra, we use the PCA data in
3--20 keV band and the HEXTE data in 20--200 keV band.

We extracted rms spectra from the PCA data using the following approach
(see also Zdziarski et al.\ 2005). First, we extracted light curves with
1/256-s resolution for the PCA absolute energy channels 0--71
(corresponding to energies from $\sim$2 to about 25--30 keV, depending
on the PCA epoch). Some of the channels were binned together to improve
statistics. Then, we calculated PDS from each
of the light curves (over 512-s intervals), subtracted the Poissonian
noise, corrected for dead-time effects (Revnivtsev, Gilfanov \&
Churazov 2000) and background (Berger \& van der Klis 1994). The
energy-dependent rms was found by integrating the PDS over the (1/512)--128
Hz frequency band.

We would like to stress the importance of background correction of the
power spectra. The fractional rms we use in this paper is the standard
deviation divided by the mean source count rate, which obviously must
exclude the background count rate (see also Berger \& van der Klis 1994).
The background variability is assumed to be Poissonian and is
subtracted from the PDS. At higher energies, $\ga$30 keV, estimating the PCA
background, which dominates most of the spectra, becomes less reliable,
so does the calculated rms. We discuss possible effects of the
high-energy background estimation is Section \ref{sec:vhs_mod}.

\section{The method}
\label{sec:method}

Fractional variability spectra are the result of the observed flux
varying differently at different energies. In order to devise any
theoretical model of energy-dependent variability, one has to make
certain assumptions about the energy spectrum. We do so by fitting the
observed spectra by a physically motivated model.

\subsection{The Comptonization model}
\label{Compton}

In this paper, we use a spectral model consisting of a soft component,
modelled by the multicolour blackbody disc emission (Mitsuda et al.\
1984), Comptonization of seed photons in the hybrid plasma ({\sc
eqpair}), and reflection of the Comptonized photons from the cold disc
(Magdziarz \& Zdziarski 1995). The model {\sc eqpair} (Coppi 1999;
Gierli{\'n}ski et al.\ 1999) calculates self-consistently microscopic
processes in a hot plasma with electron acceleration at a power-law
rate with an index $\Gamma_{\rm inj}$, in a background thermal plasma
with a Thomson optical depth of ionization electrons, $\tau$. The
electron temperature, $T$, is calculated from the balance of Compton
and Coulomb energy exchange, taking into account pair production as
well. The last two processes depend on the plasma compactness,
$\ell\equiv {\cal L}\sigma_{\rm T}/({\cal R} m_{\rm e} c^3)$, where
${\cal L}$ is a power supplied to the hot plasma, ${\cal R}$ is its
characteristic size, and $\sigma_{\rm T}$ is the Thomson cross section.
We then define the hard compactness, $\ell_{\rm h}$, corresponding to
the power supplied to the electrons, and the soft compactness,
$\ell_{\rm s}$, corresponding to the power in soft seed photons
irradiating the plasma (which are assumed to be emitted by a blackbody
disk with the maximum temperature, $T_{\rm s}$). The compactness
corresponding  to the electron acceleration and to a direct heating of
the thermal electrons are denoted as $\ell_{\rm nth}$ and $\ell_{\rm
th}$, respectively, and $\ell_{\rm h} = \ell_{\rm nth} + \ell_{\rm
th}$. Details of the model are given in Gierli{\'n}ski et al.\ (1999).
The non-thermal electrons are accelerated between the Lorentz factors
$\gamma_{\rm min}$ and $\gamma_{\rm max}$, which we assume to be 1.3
and $100$, respectively.

We then use the best-fitting spectral model to build rms($E$) models
which we can compare with variability data. Here we apply two different
approaches to rms($E$) spectra: a simple `two-component' model that has
been used in previous works, and a novel variable parameter idea, based
on physics of emission of the accretion flow.

\subsection{Two-component variability}

The simplest approach to energy-dependent variability is to consider a
number of spectral components with different variability amplitudes. We
assume in this paper that the energy spectrum consists of two
components, the soft (blackbody disc) and the hard (Comptonization). We
neglect reflection variability in this model, but discuss its possible
effects in Section \ref{sec:reflection}. When the
spectral components vary with different variances they create
energy-dependent variability, as fraction of each component in the
total spectrum changes as a function of energy. The total variance at a
given energy, $E$, is
\begin{equation}
\sigma^2(E) = \sigma^2_{\rm s}(E) + \sigma^2_{\rm h}(E) + 2\sigma_{\rm sh}(E),
\end{equation}
where $\sigma^2_{\rm s}(E)$ and $\sigma^2_{\rm h}(E)$ are the soft and hard
variances and $\sigma_{\rm sh}(E)$ is their covariance. In the following
approach, we assume that the soft and hard component variabilities are fully
correlated, so $\sigma_{\rm sh}(E) = \sigma_{\rm s}(E) \sigma_{\rm h}(E)$ and
thus
\begin{equation}
\sigma(E) = \sigma_{\rm s}(E) + \sigma_{\rm h}(E).
\end{equation}
We define fractional rms variability as rms$(E) \equiv \sigma(E)/F(E)$,
where $F(E)$ is the time-averaged flux. The total fractional rms
variability is
\begin{equation}
{\rm rms}(E) = r(N_{\rm s}){F_{\rm s}(E) \over F(E)} + r(N_{\rm h}){F_{\rm
h}(E) \over F(E)}.
\end{equation}
The $F(E)$, $F_{\rm s}(E)$ and $F_{\rm h}(E)$ are taken from the
best-fitting spectral models, while $r(N_{\rm s})$ and $r(N_{\rm h})$
are free parameters. As fractional variabilities of the normalization,
$r(N_{\rm s})$ and $r(N_{\rm h})$ are energy-independent quantities.

The two- or multicomponent approach has been used for variability
studies before (e.g., Rao et al.\ 2000; Vaughan \& Fabian 2004;
Zdziarski et al.\ 2005). There are, however, obvious caveats of this
model, as it does not take spectral variability of Comptonization into
account. Clearly, where the physical properties of the accretion flow
change, we expect, in particular, the spectral slope of Comptonization
to vary as well.

\subsection{Parameter variability}

Therefore, we have created another model based on variability of
physical properties of the accretion flow. Our spectral model (Section
\ref{Compton}, consisting of the multicolour blackbody disc and
Comptonization (again, we neglect effects of reflection on
variability), is a function of several parameters: $F = F(p_1, p_2,
..., p_n, E)$. We allow for variation of a given parameter $p_i$ with
Gaussian distribution (but we test lognormal distribution as well)
around its best-fitting mean value $\overline{p_i}$ with standard
deviation $\sigma(p_i)$ [or fractional rms $r(p_i) \equiv
\sigma(p_i)/\overline{p_i}$]. This, in turn, causes the whole X-ray
spectrum to vary, generating various patterns of energy-dependent
variability. Below, we concentrate on varying soft ($\ell_{\rm s}$) and
hard ($\ell_{\rm h}$) compactness of Comptonization, which correspond
to varying luminosity or power released in the seed photons and
electrons, respectively. We also check for effects of other parameters
varying.

\section{Results}

\subsection{Hard state} \label{sec:hard_state}

\begin{figure*}
\begin{center}
\leavevmode \epsfxsize=12cm \epsfbox{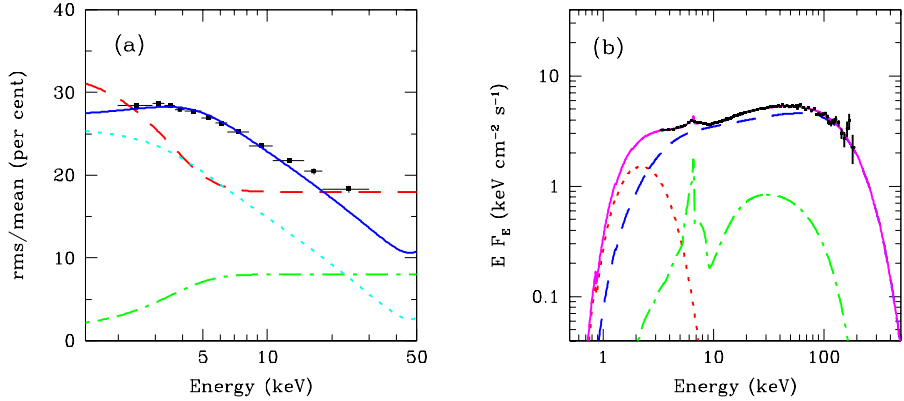}
\end{center}

\caption{(a) Hard state (obsid 60113-01-04-00) rms spectrum of XTE
J1650--500 observed by {\it RXTE}. The dashed curve (red in colour)
represents the two-component variability model, with $r(N_{\rm s}) =
36\%$ and $r(N_{\rm h}) = 18\%$, clearly not consistent with the data.
The solid (blue in colour) curve shows the model with varying soft
photon input (dotted curve, cyan in colour) at $r(\ell_{\rm s}) = 25\%$
plus an additional hard component normalization variability
(dash-dotted curve, green in colour) with $r(N_{\rm h}) = 8\%$. (b)
Energy spectrum of the same observation, with the unfolded PCA and
HEXTE data together with the best-fitting model. The model consists of
the following components: the soft component modelled by {\sc diskbb}
(dotted curve, red in colour, showing unscattered seed photons only),
its Comptonization in thermal plasma (dashed curve, blue in colour),
and reflection (dash-dotted curve, green in colour). The solid curve
(magenta in colour) shows their sum.}
\label{fig:hard_rmspec}
\end{figure*}

Fig.\ \ref{fig:hard_rmspec}(a) shows an rms spectrum of the black hole
binary XTE J1650--500 in the hard X-ray spectral state. This type of
spectrum, where fractional rms variability decreases with energy is
common in the hard state of black hole binaries. Another common
variability type in this state is a flat rms($E$), which we discuss
later in the paper. We are going to interpret the observed rms($E$)
making use of the energy spectral model fitted to the PCA/HEXTE data.
The energy spectrum of this observation with its best-fitting model is
presented in Fig.~\ref{fig:hard_rmspec}(b). It is well described
($\chi_\nu^2 = 132/129$) by purely thermal Comptonization in a plasma
with the hard-to-soft compactness ratio of \lhs\ $= 3.8^{+0.2}_{-0.1}$,
and the optical depth, $\tau = 1.82^{+0.13}_{-0.06}$. The
self-consistently computed temperature of the electrons is 47 keV, and
the soft component temperature is $kT_{\rm s} = 0.72^{+0.25}_{-0.19}$
keV. We also found Compton reflection with the amplitude of
$\Omega/2\pi = 0.18\pm 0.03$ and the ionization of the reflector of
$\lg(\xi/1\,{\rm erg\,cm\,s}^{-1}) = 3.1\pm 0.2$. This kind of a
spectrum is typical of black-hole binaries (e.g. Gierli{\'n}ski et al.\
1997; Zdziarski et al.\ 1998; Di Salvo et al.\ 2001) in the hard state
(but see our discussion of the soft component in Section
\ref{sec:hard_mod}).

We first try the two-component model, considering soft and Comptonized
components varying in luminosity, but not in the spectral shape. When
the energy spectrum is dominated by one component, the amplitude of
variability does not depend on energy. A spectrum varying in
normalization only yields the same {\em fractional} variability
amplitude at all energies, regardless of its spectral shape. Therefore,
at energies $\ga$5 keV, where Comptonization dominates, the predicted
rms spectrum is flat. This behaviour is not consistent with the
observed rms($E$) relation in the hard state of XTE J1650--500, where
the rms amplitude decreases with energy. In
Fig.~\ref{fig:hard_rmspec}(a) we show a model of two-component
variability pattern (dashed curve) compared to the data. Clearly, this
kind of variability is not consistent with observations. Including
reflection as a third independently varying component cannot resolve
this problem either.

Then, to explain the decreasing rms($E$) in the hard state, we
considered variability of the seed photon input (Z02). We allowed for
variations in the soft component normalization, $N_{\rm s}$ (but not
in its temperature). We assumed that the seed photons for Comptonization
originated from the soft spectral component. Therefore, $\ell_{\rm s}$
was linearly proportional to the soft component luminosity, so it
varied with the same fractional rms as $N_{\rm s}$. The hard
compactness, $\ell_{\rm h}$, was constant, so the hard-to-soft
compactness ratio, $\ell_{\rm h}/\ell_{\rm s}$, varied as well. The
hard-to-soft compactness ratio sets the energy balance between the seed
photons and hot electrons, and is directly responsible for the hardness
(spectral index) of the energy spectrum (e.g., Coppi 1999). Therefore,
varying $\ell_{\rm s}$ caused pivoting of the whole spectrum
around $\sim$50 keV (as shown in Fig. \ref{fig:3models}a). We assumed
$\overline{\ell_{\rm s}} = 10$ and $r(\ell_{\rm s}) = 25\%$. As a
result, the fractional rms decreased with energy, as showed by the
dotted (cyan in colour) curve in Fig.~\ref{fig:hard_rmspec}(a). This
model was well below the observed rms, but simply increasing $\ell_{\rm
s}$ variability did not help because at a higher $r(\ell_{\rm s})$ the
model rms($E$) became much steeper than the observed one. To resolve
the problem we allowed for additional variability of the hard component
luminosity (or normalization), with $r(N_{\rm h}) = 8\%$. The resulting
rms($E$) matched the data well. We note that we have not performed
formal fits of our models to the observed rms spectra. We have only
found a model that visually matched the observed rms pattern. This
best-matching model can be interpreted as strongly varying seed photon
input (resulting in the pivoting hard component) accompanied by a much
weaker variation in the hard component luminosity. We have tested other
model parameters variability, but none of them produced rms($E$) even
remotely consistent with the data.

We also tested the effect of a probability distribution of varying
$\ell_{\rm s}$ different then Gaussian. A feasible alternative is
to use lognormal distribution, as it has been found to fit the
distribution of fluxes from Cyg X-1 (Uttley, McHardy \& Vaughan 2005).
We have repeated the above calculations for a lognormal distribution of
$\ell_{\rm s}$ and found only negligible difference between the
lognormal and Gaussian distribution results.

\subsection{Soft state} \label{sec:soft_state}

\begin{figure*}
\begin{center}
\leavevmode \epsfxsize=12cm \epsfbox{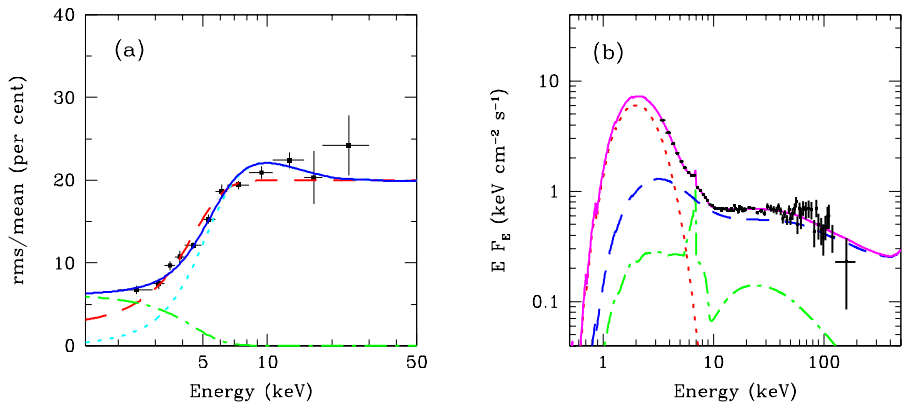}
\end{center}

\caption{(a) Soft state (obsid 60113-01-18-01) rms spectrum of XTE
J1650--500. The dashed curve (red in colour) represents the
two-component variability model, with $r(N_{\rm s}) = 0\%$ and
$r(N_{\rm h}) = 20\%$. The solid curve (blue in colour) shows the model
with varying hard compactness (dotted curve, cyan in colour) at
$r(\ell_{\rm h}) = 17\%$ plus additional soft component normalization
variability (dash-dotted curve, green in colour) with $r(N_{\rm s}) =
7\%$. (b) The energy spectrum of the same observation. The model
components are identical to those in Fig.~\ref{fig:hard_rmspec}(b),
except for Comptonization taking place in the hybrid,
thermal/non-thermal plasma.}

\label{fig:soft_rmspec}
\end{figure*}

Fig.\ \ref{fig:soft_rmspec}(a) shows an rms spectrum of XTE J1650--500
in the soft X-ray spectral state. The pattern of variability is
distinctly different from the hard state in
Fig.~\ref{fig:hard_rmspec}(a). It rises with energy and saturates above
$\sim$10 keV. We have looked through all rms spectra of XTE J1650--500
and found that this pattern is characteristic for all the soft state
spectra.

The soft-state energy spectrum (Fig.~\ref{fig:soft_rmspec}b) is well
described ($\chi_\nu^2 = 115/117$) by a hybrid Comptonization where
electrons are injected to the plasma with a power-law distribution with
the index fixed at $\Gamma_{\rm inj} = 2.5$. The fraction of
non-thermal power transferred to the electrons (as opposed to thermal
heating) is \lnth\ $= 0.84_{-0.10}^{+0.03}$. The best-fitting
hard-to-soft compactness ratio is \lhs\ $= 0.38_{-0.15}^{+0.24}$ and the
optical depth is $\tau = 1.2^{+0.6}_{-0.4}$. The self-consistently
computed temperature of the electrons is 11 keV. The disc temperature
is $kT_{\rm s} = 0.54_{-0.03}^{+0.02}$ keV. We also found Compton
reflection with the amplitude of $\Omega/2\pi = 0.22^{+0.04}_{-0.12}$
and the reflector ionization of $\lg(\xi/1\,{\rm erg\,cm\,s}^{-1}) =
4.7_{-0.8}^{+0.3}$. This kind of a spectrum is typical of black hole
binaries (e.g. Gierli{\'n}ski et al.\ 1999; Frontera et al.\ 2001) in
the soft state.

Contrary to the hard state (Section \ref{sec:hard_state}), the rms
spectrum is very well described by the two-component model, where disc
and Comptonization are allowed to vary in normalization only. We show
this model, with dominant variability of Comptonization, in
Fig.~\ref{fig:soft_rmspec}(a), with the dashed (red in colour) curve. A
characteristic feature of this model is a break at around 7 keV, above
which Comptonization dominates the spectrum (see
Fig.~\ref{fig:soft_rmspec}b) and rms($E$) becomes flat. The energy of
this break is directly related to the disc temperature and we found
that rms($E$) saturates at $\sim$15 $kT_{\rm s}$.

Another model that can explain rms increasing with energy is variable
hard power input (Z02). This is opposite to the soft photon input
variability in the hard state: this time we kept $\ell_{\rm s}$
constant and allowed $\ell_{\rm h}$ to vary. This caused the energy
spectrum to vary in a way depicted in Section \ref{sec:patterns} (see
Fig.\ \ref{fig:3models}(b) below). The result for $r(\ell_{\rm h}) =
17\%$ is shown by the dotted (cyan in colour) curve in
Fig.~\ref{fig:soft_rmspec}(a). The rms($E$) pattern predicted by this
model matches the data well, except for the deficiency in rms below
$\sim$4 keV. We have accounted for this deficiency by adding little
variability in the disc. The total rms spectrum (solid curve) matches
the data well, and is similar in shape to the two-component model. The
energy of the break at $\sim$10 keV is also directly related to the
seed photon temperature. As in the hard state, the lognormal
distribution of $\ell_{\rm h}$ produced very similar results to the
Gaussian distribution.

\subsection{Very high state} \label{sec:vhs}

\begin{figure*}
\begin{center}
\leavevmode \epsfxsize=12cm \epsfbox{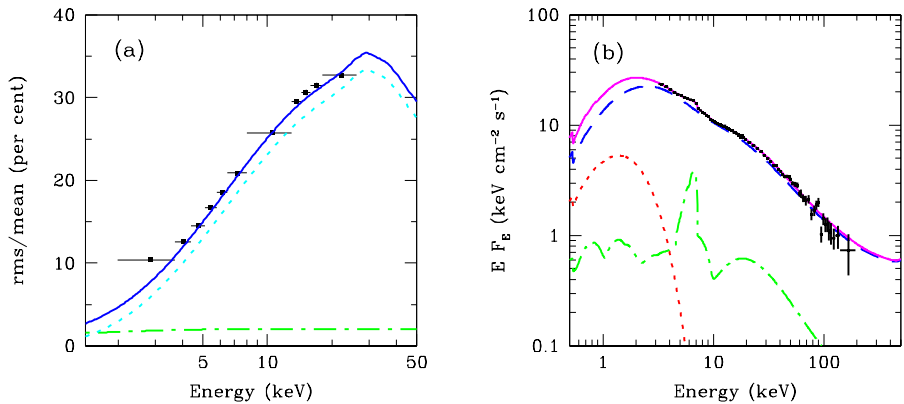}
\end{center}

\caption{(a) Very high state (obsid 30191-01-01-00) rms spectrum of XTE
J1550-564. The solid curve (blue in colour) represents the model of
variability consisting of varying hard compactness (dotted curve, cyan
in colour) with $r(\ell_{\rm h}) = 17\%$ plus additional hard component
normalization variability (dash-dotted curve, green in colour) with
$r(N_{\rm h}) = 2\%$. (b) The energy spectrum of the same observations.
The model components are identical to those in
Fig.~\ref{fig:hard_rmspec}(b), except for Comptonization taking place
in the hybrid, thermal/non-thermal plasma.}

\label{fig:vhs_rmspec}
\end{figure*}

Fig.~\ref{fig:vhs_rmspec}(a) shows rms spectrum of another black hole
XTE J1550--564 in the very high spectral state. Like in the soft-state
spectrum in Fig.~\ref{fig:soft_rmspec}(a), the fractional variability
increases with energy, though there is neither break nor saturation up
to at least $\sim$20 keV.

The very high state energy spectrum of XTE J1550--564
(Fig.~\ref{fig:vhs_rmspec}b) is well described by hybrid Comptonization
(see also Gierli{\'n}ski \& Done 2003) with weak apparent contribution
from unscattered disc photons. The best-fitting model ($\chi_\nu^2 =
117/122$) parameters are: hard-to-soft compactness ratio, \lhs\
$=0.80_{-0.03}^{+0.08}$, non-thermal fraction, \lnth\
$=0.83_{-0.15}^{+0.06}$, electron injection power-law index of
$\Gamma_{\rm inj} = 3.3_{-0.2}^{+0.1}$, and the optical depth of $\tau
= 4.73_{-0.10}^{+0.35}$. The electron temperature is 4 keV, and the
disc temperature is $kT_{\rm s} = 0.52_{-0.09}^{+0.06}$ keV. The
reflection amplitude is $\Omega/2\pi = 0.20_{-0.11}^{+0.16}$ and its
ionization, $\lg(\xi/1\,{\rm erg\,cm\,s}^{-1}) = 3.6_{-1.1}^{+0.2}$.

The two-component variability that has successfully described the
soft-state rms spectrum cannot explain the very high state data, as it
predicts a break at $\sim$6~keV and flat rms($E$) above the break
(similar to the model shown in Fig.~\ref{fig:soft_rmspec}a). The
flattening of rms($E$) occurs always at energies where Comptonization
dominates and does not depend on its spectral shape. On the other hand,
the $\ell_{\rm h}$ variability model matched the data very well, though
some additional hard component variability contribution was required,
similarly to the hard state. Interestingly, in the soft state, the
$\ell_{\rm h}$ variability created rms($E$) pattern with a break at
$\sim 15 kT_{\rm s}$, which is not seen here until about 30 keV. And
again, using the lognormal distribution of $\ell_{\rm h}$ had very little
effect on our results.

\section{Patterns of variability}
\label{sec:patterns}

\begin{figure*}
\begin{center}
\leavevmode \epsfxsize=14cm \epsfbox{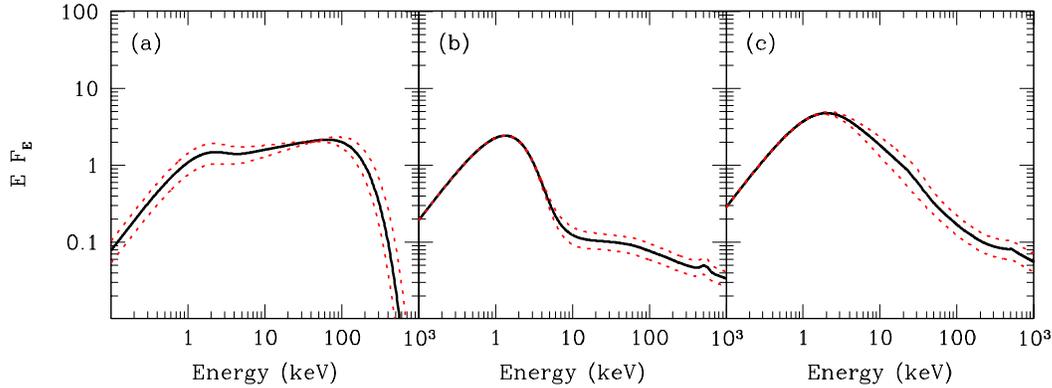}
\end{center}

\caption{Visualization of model spectral variability (dotted curves,
red in colour) with respect to the best-fitting spectrum in a given
state (solid black curve). (a) Hard state: variability of the soft
photon input, $\ell_{\rm s}$. (b) Soft state and (c) very high state:
variability of the hard power, $\ell_{\rm h}$.}

\label{fig:3models}
\end{figure*}

\begin{figure*}
\begin{center}
\leavevmode \epsfxsize=17cm \epsfbox{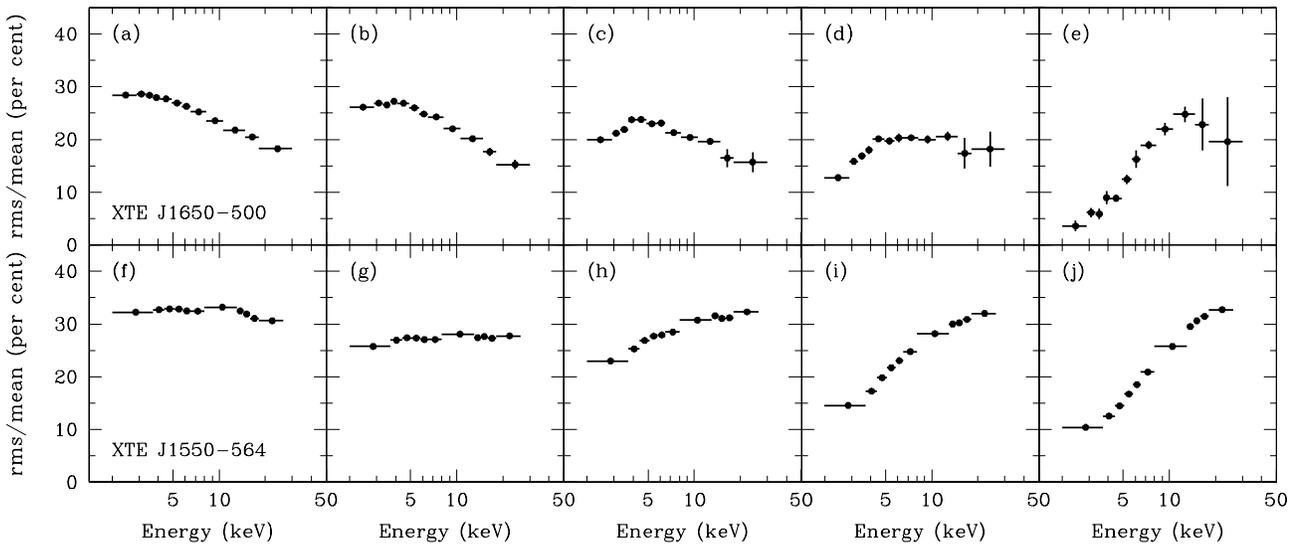}
\end{center}

\caption{Characteristic patterns of rms($E$) variability observed from two
Galactic black hole binaries. The upper row shows the evolution XTE
J1650--500 during its outburst and transition from the hard (a) through
the intermediate (b-c) to the soft (d-e) spectral state. Spectra from
observations 60113-01-X are shown, where X is: (a) 04-00, (b) 08-00, (c)
12-00, (d) 13-01 and (e) 24-00. The lower row shows the evolution of XTE
J1550--564 in the beginning of its 1998 outburst and transition from the
hard to the very high state. Spectra from observations 30188-06-X are
shown, where X is: (f) 01-00, (g) 01-02, (h) 04-00, (i) 06-00, and (j) is
from 30191-01-01-00.}

\label{fig:trans}
\end{figure*}

\begin{figure*}
\begin{center}
\leavevmode \epsfxsize=17cm \epsfbox{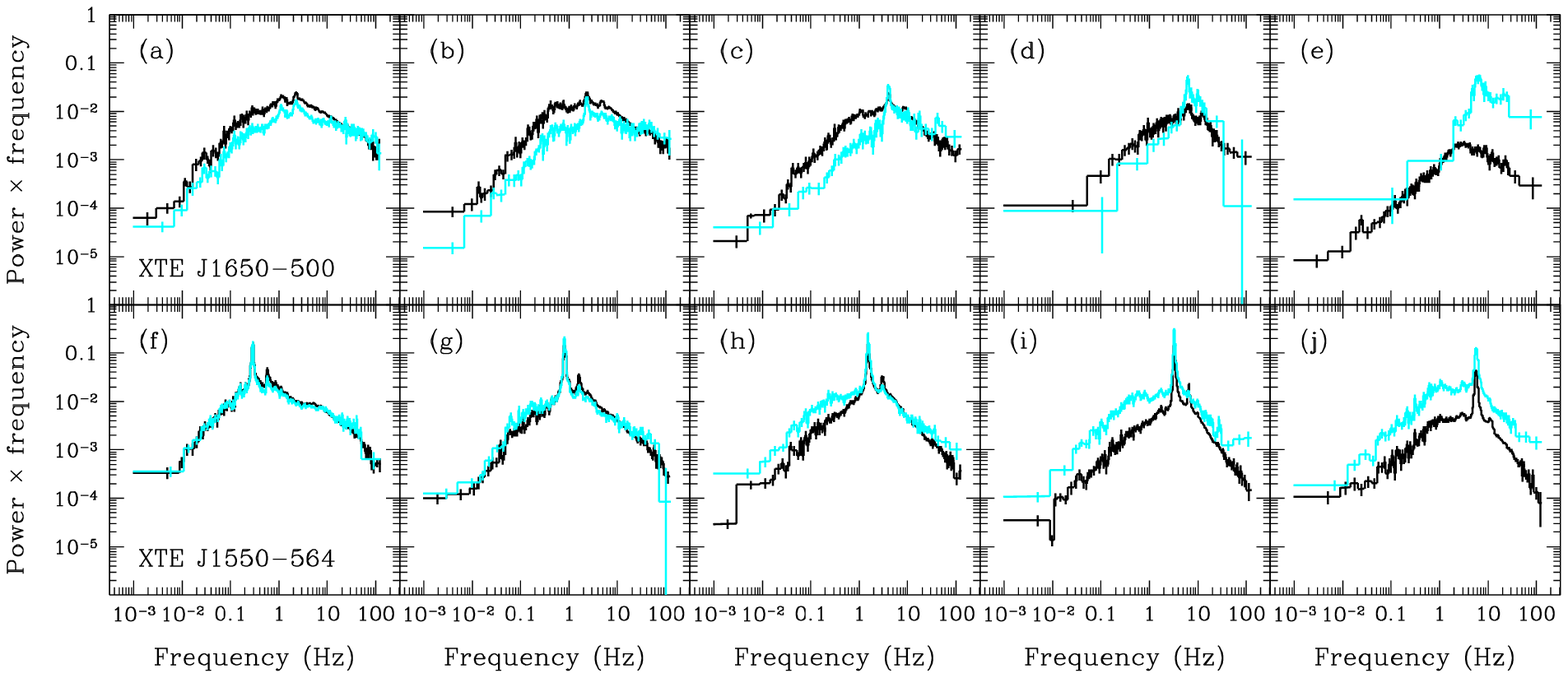}
\end{center}

\caption{The power density spectra corresponding to the rms($E$)
dependencies  shown in Fig.~\ref{fig:trans}. The low- and high-energy
PDS are shown in black and grey (cyan in colour), respectively. The
low-energy band corresponds to the PCA absolute channels 0--25 (XTE
J1650--500) and 0--35 (XTE J1550--564). The high-energy band
corresponds to the channels above those up to the channel 71. The
energies corresponding to these two bands are $\sim$2 to $\sim$13 and
$\sim$13 to $\sim$25 keV, respectively. The spectra have been rebinned
for clarity. Note the maximum of the frequency times power shown here
corresponds to the maximum of the variability power per log of
frequency. }

\label{fig:pds}
\end{figure*}

In previous section, we have presented characteristic rms spectra in
three X-ray spectral states and possible models matching these spectra.
Certainly, the two-component model where the disc and Comptonization
vary independently is only a zeroth-order approximation. Except for a
specific case described in Section \ref{sec:geometry} below, there
should be a feedback: a change in the disc luminosity changes the
supply of the seed photons for Comptonization, which in turn should
affect the spectral shape of the Comptonized component. Therefore, the
models based on varying a parameter in a physical model where the disc
and Comptonization are linked together should provide better physical
picture of variability.

In Fig.~\ref{fig:3models}, we visualize model spectral variability in
each spectral state, with varying soft photon input and hard
compactness. For simplicity, we do not include additional variability
components required by the data (see Figs. \ref{fig:hard_rmspec},
\ref{fig:soft_rmspec} and \ref{fig:vhs_rmspec}).

In Fig.~\ref{fig:trans}, we show several characteristic rms spectra
from XTE J1650--500 and XTE J1550--564 covering hard, soft and very
high spectral states. We also show the corresponding power density
spectra in Fig.~\ref{fig:pds}. These PDS were extracted over the same
frequency band [(1/512)-128 Hz] as those used for creation of the rms
spectra. To illustrate energy dependence we show the low- ($\la$13 keV)
and high-energy ($\ga$13 keV) power spectra. Clearly, in many cases not
only the normalization, but also the PDS shape changes as a function of
energy.

Below, we discuss in details possible variability models in each of the
spectral states shown in Figs.~\ref{fig:trans}--\ref{fig:pds} and
give their theoretical interpretation.

\subsection{Hard state}
\label{sec:hard_mod}

\begin{figure}
\begin{center}
\leavevmode \epsfxsize=8.5cm \epsfbox{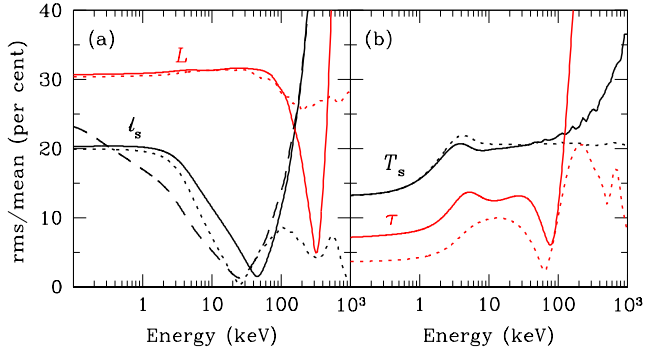}
\end{center}

\caption{Models of rms($E$) variability in the hard state. (a)
Variability of soft photon input, $r(\ell_{\rm s}) = 20\%$, and of the total
luminosity, $r(L) = 30\%$, assuming $\tau \propto L^{2/7}$.
(b) Variability of seed photon temperature, $r(T_{\rm s}) = 5\%$, and of
optical depth, $r(\tau) = 50\%$. Solid curves correspond to the
best-fitting thermal model from Section \ref{sec:hard_state}, dotted
curves correspond to the same model, but with non-thermal fraction set
to 1. The dashed curve in panel (a) represents $\ell_{\rm s}$
variability of an alternative (thermal) model with the seed photon
temperature set to 0.1 keV.}
\label{fig:hard_mod}
\end{figure}

In the hard state, the rms spectrum is either flat
(Fig.~\ref{fig:trans}f,g) or smoothly decreasing with energy
(Fig.~\ref{fig:trans}a,b). The flat rms($E$) simply corresponds to a
situation where the entire spectrum (or Comptonization only, when the
disc is not visible in the observed bandwidth) varies in normalization
(luminosity) but not in spectral shape (see also Z02). If we assume
that luminosity variations are due to changes in the accretion rate,
than the only spectral effect we might expect is weak variation in the
shape of the high-energy cutoff $\ga$100 keV, as the optical depth
varies following variations in the flow density. We have considered a
particular model of the advection dominated flow, where $\tau \propto
L^{2/7}$ (Zdziarski 1998). The result is shown in
Fig.~\ref{fig:hard_mod}(a) with a solid grey (red in colour) curve. The
minimum at $\sim$300 keV is due to fortuitous intersection of the
spectra resulting from this variability prescription at that energy.
Additionally, we have checked the result of replacing the thermal
Comptonization by that with fully non-thermal injection ($\ell_{\rm
nth}/\ell_{\rm h}=1$), which turns out to be negligible at $\la$100 keV
(dotted curve).

The other pattern observed in the hard state is the rms smoothly
decreasing with energy. Z02 explained a similar pattern observed on
much longer timescales by variable seed photon input, $\ell_{\rm s}$.
We found this solution consistent with our observations. The result of
this model is plotted in Fig.~\ref{fig:hard_mod}(a) with a solid black
curve. The characteristic feature of this type of variability is
pivoting of the spectrum around $\sim$20--50 keV (Fig.
\ref{fig:3models}a). Because of that, the rms($E$) reaches a very deep
minimum around this energy, so the decline in rms above $\sim$5 keV is
very steep, and steeper than that observed (see
Fig.~\ref{fig:hard_rmspec}a). This can be understood when we notice
that in our model we assumed variations in $\ell_{\rm s}$ only, with
Comptonizing plasma simply responding to these changes in the seed
photons, while the data required some variation in the power released
in the Comptonized component, $\ell_{\rm h}$. The full treatment of
this problem within our model would require two-dimensional variation
in $\ell_{\rm s}$ and $\ell_{\rm h}$, with some particular relation
between the two parameters assumed. For the sake of simplicity, in
Section \ref{sec:hard_state} we allowed for variations in just one
parameter ($\ell_{\rm s}$) and added a separate rms($E$) component
corresponding to variations in the hard component luminosity. This may
correspond to an intermediate case between the flat-rms constant
spectral shape model and pure seed photon input variations.

The dotted curves in Fig.~\ref{fig:hard_mod} show the effect of
non-thermal acceleration in the hot plasma. Because it is dominated by
the Compton cooling (and often termed as a photon-starved plasma), the
electrons in the hard state are efficiently thermalized even when the
power provided to them is entirely in the form of non-thermal
acceleration (Coppi 1999; Zdziarski, Coppi \& Lamb 1990). The effect on
the energy spectrum is visible only at high energies and difficult to
measure (McConnell et al.\ 2002). Our PCA/HEXTE spectrum from Section
\ref{sec:hard_state} can be fitted by Comptonization with thermal and
non-thermal electron injection equally well. The rms($E$) of the
thermal spectrum tends to rise steeply at around high-energy cutoff.
This is because the spectrum around the cutoff varies in the direction
perpendicular to the curve representing the spectrum
(Fig.~\ref{fig:3models}a), due to changes in the electron temperature
as it adjusts itself to satisfy the energy balance (Z02). However, when
a slight non-thermal tail added to the spectrum, the rms does not
increase that dramatically and the high-energy variability pattern
resembles that one in the soft state (Fig.~\ref{fig:3models}b).

We have also investigated rms spectra resulting from varying seed
photon temperature and optical depth of the Comptonizing plasma. The
results (Fig.~\ref{fig:hard_mod}b) are highly inconsistent with
observations, restricting our models to the variability in $\ell_{\rm
s}$ and luminosity, as discussed above.

An interesting feature of the $\ell_{\rm s}$ variability in the hard
state is its dependence on the seed photon temperature. The rms becomes
constant below a certain energy, directly related to the temperature of
the seed photons. In our best-fitting model from Section
\ref{sec:hard_state}, the seed photons originate from the soft
component of temperature of $\sim$0.7 keV, and the rms($E$) flattens
below $\sim$2 keV (Fig.~\ref{fig:3models}a). We would like to point out
that the soft component in this particular model {\em cannot} be a
standard Shakura-Sunyaev disc (Shakura \& Sunyaev 1973), as it is too
hot ($\sim$0.7 keV) and too small [$R_{\rm in} \approx 18$ km for a
distance of 4 kpc (Tomsick et al.\ 2003) and inclination of $30\degr$
(S{\'a}nchez-Fern{\'a}ndez et al.\ 2002)] for a disc. Instead, we
probably see the so-called soft excess, an additional thermal
Comptonization component hotter than the disc (see, e.g., Di Salvo et
al.\ 2001; Frontera et al.\ 2001). The dashed curve in
Fig.~\ref{fig:hard_mod}(a) shows an alternative model in which the seed
photon temperature was 0.1 keV and the soft excess was fitted as an
additional component (which variability was not modelled). Clearly, the
slope of rms($E$) below $\sim$2 keV is distinctly different from the
model in which the seed photons came from the observed soft excess. It
is arguable whether the flattening of rms($E$) a low energies is
present in the hard-state data (Fig.~\ref{fig:hard_rmspec}a) and
additional data from instruments sensitive below 1 keV (e.g., {\it
XMM-Newton\/} or {\it Chandra}) are required to confirm its veracity.
If flattening is real then the seed photons are not from the disc (or
at least not entirely from the disc) but from the soft excess. This can
yield crucial constraints on the origin of the soft excess and geometry
of the accretion flow.

\subsection{Soft state}
\label{sec:soft_mod}

\begin{figure*}
\begin{center}
\leavevmode \epsfxsize=17.5cm \epsfbox{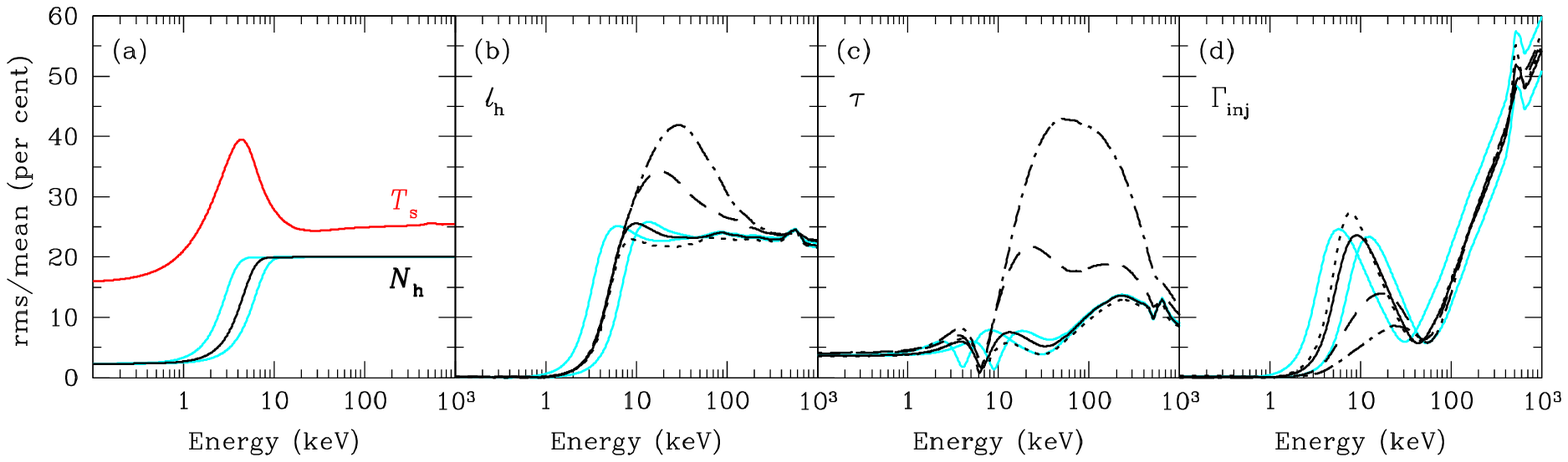}
\end{center}

\caption{Models of rms($E$) variability in the soft state. (a)
Variability of the hard component normalization, $r(N_{\rm h}) = 20\%$
and of the seed photon temperature, $r(T_{\rm s}) = 6\%$. (b)
Variability of the hard compactness, $r(\ell_{\rm h}) = 20\%$,
(c) the optical depth, $r(\tau) = 50\%$, and (d) the electron injection index,
$r(\Gamma_{\rm inj}) = 20\%$. Solid curves correspond to the
best-fitting hybrid model from Section \ref{sec:soft_state}. Grey (cyan
in colour) curves on each side of the solid curve show the same model,
but with seed photon temperature altered by $\pm0.2$ keV with respect
to the best fit. Dotted, dashed and dot-dashed black curves correspond
to \lnth\ = 1.0, 0.4 and 0.2, respectively.}

\label{fig:soft_mod}
\end{figure*}

\begin{figure*}
\begin{center}
\leavevmode \epsfxsize=17.5cm \epsfbox{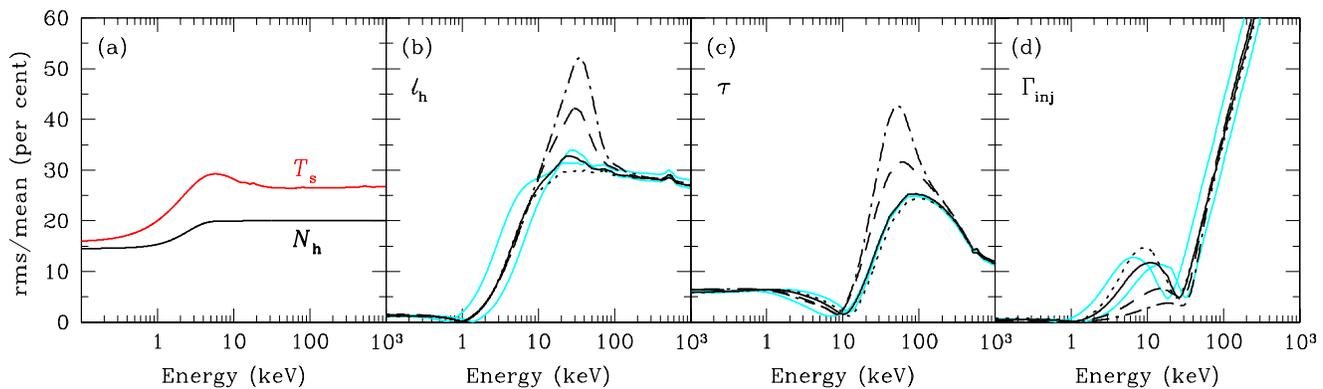}
\end{center}
\caption{Models of rms($E$) variability in the very high state. (a)
Variability of the hard component normalization, $r(N_{\rm h}) = 20\%$,
and of the seed photon temperature, $r(T_{\rm s}) = 6\%$. (b)
Variability of the hard power, $r(\ell_{\rm h}) = 20\%$,
(c) the optical depth, $r(\tau) = 20\%$, and (d) the electron injection index,
$r(\Gamma_{\rm inj}) = 20\%$. Solid curves correspond to the
best-fitting hybrid model from Section \ref{sec:vhs}. Grey (cyan in
colour) curves on each side of the solid curve show the same model, but
with seed photon temperature altered by $\pm 0.2$ keV with respect to
the best fit. Dotted, dashed and dot-dashed black curves correspond to
\lnth = 1.0, 0.4 and 0.2, respectively.}
\label{fig:vhs_mod}
\end{figure*}

Soft-state rms spectra are distinctly different from hard state ones.
While the hard state typically showed either flat rms or a decrease in
power with energy, rms($E$) increases and then saturates at higher
energies in the soft state (see also Zdziarski et al.\ 2005). Similar
spectral changes have been also reported in variability of QPOs (e.g.\
Rodriguez et al.\ 2004a). The panels (b), (c) and (d) in
Fig.~\ref{fig:trans} show transition from the hard to the soft state.
An interesting feature of these spectra is the break at $\sim$2--4 keV,
present throughout the transition.

Using our best-fitting non-thermal spectral model, we have found in
Section \ref{sec:soft_state} two similar rms($E$) models that match the
observed rms spectrum. The first one assumed variability in the hard
component normalization, $N_{\rm h}$, only. It produces a quick
increase in rms and saturation above a few keV, as shown in
Fig.~\ref{fig:soft_mod}(a) with the black curve. The grey (cyan in
colour) curves show the effect of different $T_{\rm s}$.

The second model involves constant $\ell_{\rm s}$ and variable
$\ell_{\rm h}$. Unlike the previous model, it takes into account
spectral response of Comptonization to the changing ratio of the power
released in the corona to that in the disc. On the other hand, changes
in the spectral shape of Comptonization (at least in the non-thermal
case) are rather small, so the rms spectrum produced by the $\ell_{\rm
h}$ variability is similar to the one from $N_{\rm h}$ variability: a
quick rise and saturation at higher energies (Fig.\ \ref{fig:soft_mod}b).

Our models do not always predict a flat rms above the break energy. In
Fig.~\ref{fig:soft_mod}(b), we also show (dashed and dot-dashed curves)
the effect of decreasing fraction of the non-thermal acceleration in the
total power, \lnth. As the thermal heating becomes more important, a
peak around 20--30 keV in rms($E$) is created. We discuss the origin of
this peak in the next section.

A common property of both models is a break in the spectrum followed by
flat rms($E$) at higher energies. The formation of the break and flat
rms can be seen in the XTE J1650--500 data following its evolution from
the intermediate to soft state in Figs.~\ref{fig:trans}(b--d). As
mentioned above, the flat rms($E$) is due to a spectral component
varying in normalization but not in shape. The break corresponds to an
energy in the spectrum above which Comptonization dominates, which is
roughly at 15 $kT_{\rm s}$ (the best-fitting seed photons temperature
was 0.54 keV).

As in the hard state, variability of either the seed photon temperature,
$T_{\rm s}$, or the optical depth of the hot plasma, $\tau$, do not provide
rms($E$) patterns consistent with the data (Fig.~\ref{fig:soft_mod}a,
c).

Z02 considered variability in the hardness of the power-law electrons
injected into the hot plasma. This can be done in two ways, either by
varying the injection index, $\Gamma_{\rm inj}$, or by varying the
high-energy cutoff in the electron distribution represented by the
maximum Lorentz factor, $\gamma_{\rm max}$. Z02 concluded that the
first pattern was inconsistent with the colour-colour and
colour-luminosity correlations, while the last one provided a good fit,
though they have not analysed rms spectra emerging from these patterns.
Here we calculate the exact form of rms($E$) both for varying
$\Gamma_{\rm inj}$ and $\gamma_{\rm max}$. We find that both of them
create very similar rms spectra, and we show one of them ($\Gamma_{\rm
inj}$) in Fig.~\ref{fig:soft_mod}(d). Both patterns are characterized
by a strong rms peak around $\sim$10 keV and are clearly inconsistent
with any rms spectrum found by us so far. Therefore, we conclude that
neither $\Gamma_{\rm inj}$ nor $\gamma_{\rm max}$ variability can
explain the observed rms($E$) patterns.

\subsection{Very high state}
\label{sec:vhs_mod}

The very high state energy spectrum is dominated by a strong
non-thermal Comptonized tail (Fig.~\ref{fig:vhs_rmspec}b; see also
Gierli{\'n}ski \& Done 2003). The relative contribution from the disc
is much weaker than in the soft state and the Comptonized tail is much
softer than in the hard state. The rms spectra are also distinctly
different. Panels (f--j) in Fig.~\ref{fig:trans} show evolution of the
rms from the hard to the very high state. The initially flat
rms($E$) becomes very steep, increasing with energy without any
apparent break up to at least $\sim$20 keV.

We have investigated the same patterns of variability as in the soft
state, using the best-fitting model from Section \ref{sec:vhs}. The
results are shown in Fig.~\ref{fig:vhs_mod}. The simplest model of
$N_{\rm h}$ variability (or, more general, two-component variability)
does not work here. As the entire energy spectrum is dominated by
Comptonization, the resulting rms($E$) is almost flat in this model,
with a slight depression below $\sim$3 keV. The observed strongly
increasing rms($E$) requires the Comptonized tail to vary in spectral
shape, not only in normalization. In Section \ref{sec:vhs}, we have
found that $\ell_{\rm h}$ variability matches the data well.
Fig.~\ref{fig:vhs_mod}(b) now shows the dependence of $\ell_{\rm h}$
variability models on the seed photon temperature and non-thermal
fraction.

\begin{figure}
\begin{center}
\leavevmode \epsfxsize=7cm \epsfbox{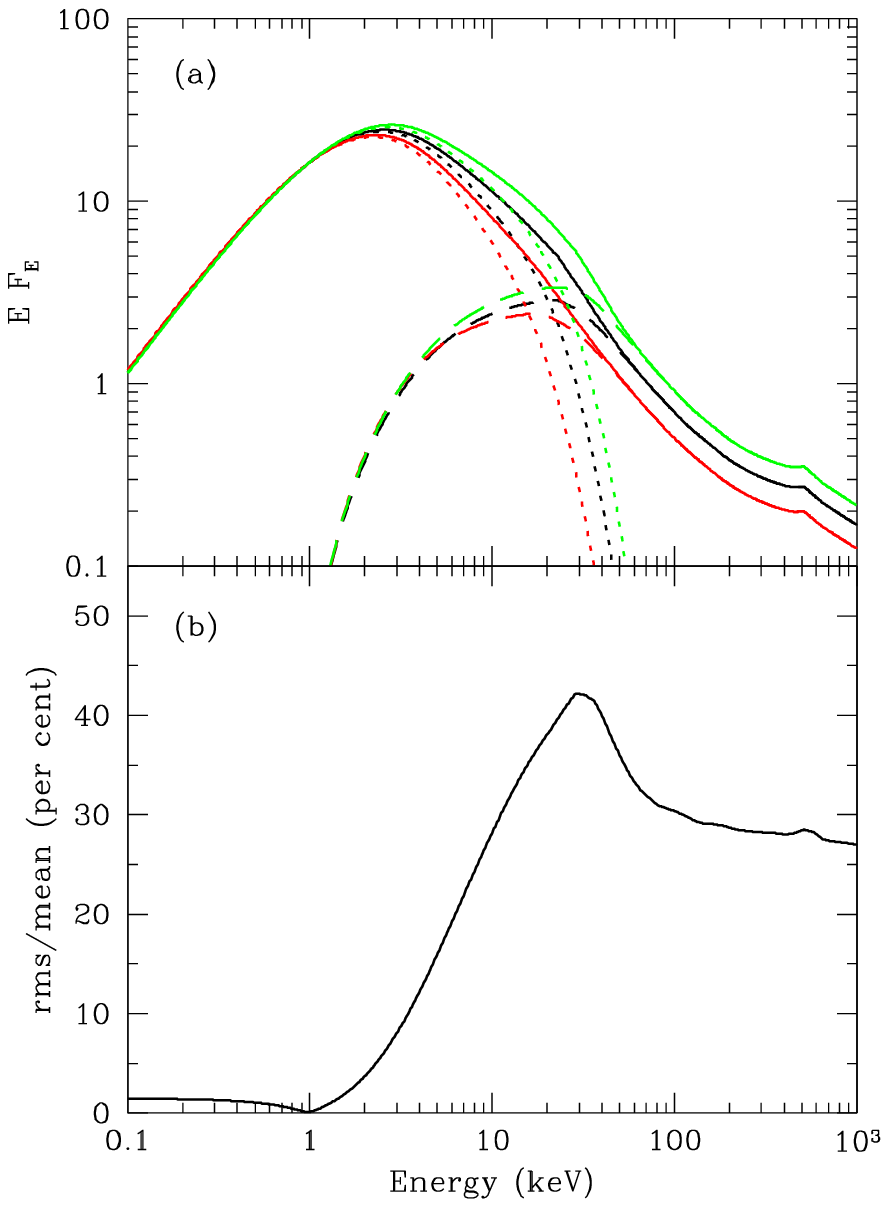}
\end{center}
\caption{(a) Visualization of model spectral variability in the very
high state due to changes of $\ell_{\rm h}$. The model shown corresponds to the
non-thermal fraction \lnth\ = 0.4. The three spectra correspond to
variation in $\ell_{\rm h}$ by $\pm 20$ per cent (top and bottom) and the
original spectrum (middle). The dotted and dashed curves represent
thermal and non-thermal components of the hybrid Comptonization. (b) The
corresponding rms spectrum. The peak at $\sim$30 keV is due to the
variability of the high-energy cutoff of the thermal-Compton component.}
\label{fig:vhs_nth}
\end{figure}

A common feature of $\ell_{\rm h}$ variability patterns in the soft and
very high states is formation of a peak in the rms spectrum at about
20--30 keV when thermal heating is present in the Comptonizing plasma,
i.e., when \lnth\ is less then 1. The origin of the peak can be
understood from the decomposition of the hybrid Comptonization spectrum
into thermal and non-thermal components (following the method of
Hannikainen et al.\ 2005). This decomposition is shown in
Fig.~\ref{fig:vhs_nth}(a) for the very high state spectrum where the
non-thermal fraction was set to \lnth\ = 0.4, corresponding to
rms($E$) shown in Fig.~\ref{fig:vhs_nth}(b). The peak in rms($E$) is
due to variations in the high-energy cutoff of the thermal component. A
similar kind of variability can be seen in the hard state
(Fig.~\ref{fig:3models}a) and it causes a dramatic increase in the
rms($E$), as seen in Fig.~\ref{fig:hard_mod}(a). In the case of the
very high state, though, the increase is suppressed at higher energies,
where the non-thermal component begins to dominate. With increasing
contribution of thermal heating, the peak becomes stronger. This is an
important feature of hybrid Comptonization and it will be interesting
to see whether it is detected in future observations.

\begin{figure}
\begin{center}
\leavevmode \epsfxsize=5cm \epsfbox{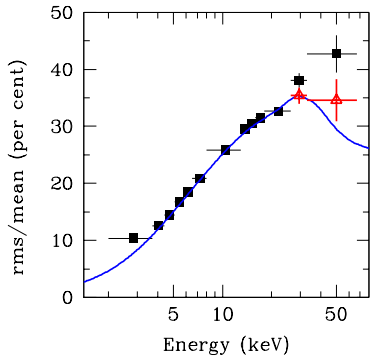}
\end{center}
\caption{The rms spectrum and model of the very high state from Section
\ref{sec:vhs} (see Fig.~\ref{fig:vhs_rmspec}a). Here, we have
extended the energy  scale to the limits of PCA capabilities. The
squares show the direct extension of the spectrum, calculated in the
standard way, as described in Section \ref{sec:data}, which is clearly
incompatible with our best model. The triangles show the rms corrected
for the background excess, estimated from the broad-band PCA/HEXTE
energy spectrum.} \label{fig:vhs_bkg}
\end{figure}

The very high state rms spectrum of XTE J1550--564 shown in
Fig.~\ref{fig:vhs_rmspec}(a) rises with energy until about 25 keV,
i.e.\ until the last PCA energy channel we use in this paper. In
Fig.~\ref{fig:vhs_bkg}, we show the same spectrum computed up to energy
of 67 keV. The last two data points (filled boxes) correspond to the
rebinned absolute PCA data channels 72--89 and 90--174. There is a
clear increasing trend in the rms, not consistent with the $\ell_{\rm
h}$ variability model. However, one should treat the PCA data above
$\sim$30 keV, where background contribution becomes important, with
great caution. To test the accuracy of background estimate we have
looked into our PCA/HEXTE energy spectrum with its best-fitting model
from Section \ref{sec:vhs}. We extended the PCA energy spectrum up to
about 67 keV and compared additional energy channels with the
best-fitting model to our standard PCA/HEXTE spectrum. It occurred that
the PCA flux in additional high-energy channels was significantly lower
than in the model, most likely due to the overestimated PCA background.
The model-to-data ratio in the energy bins corresponding to the two
additional points in Fig.~\ref{fig:vhs_bkg} was 0.93 and 0.82. We have
introduced these corrections to the mean count rate in computed
fractional rms (which is rms/mean). The corrected data points are shown
in Fig.~\ref{fig:vhs_bkg} by open triangles. Now they appear to be
consistent with the $\ell_{\rm h}$ variability model. We would like to
stress that by doing this we have pushed the PCA data to the limits
where systematic uncertainties are not very well known. Therefore this
result should be treated with caution.

Finally, we have calculated model variability patterns for varying
$T_{\rm s}$, $\tau$ and $\Gamma_{\rm inj}$ (Fig.~\ref{fig:vhs_mod}a, c,
d). As it was the case in the soft state, we found them not consistent
with the data.

\section{Accretion flow geometry}
\label{sec:geometry}

Z02 considered a truncated disc geometry to explain the observed
rms($E$) patterns. Within this model, in the hard state the accretion
disc is truncated at some radius and replaced by a hot optically thin
inner flow (e.g.\ Poutanen, Krolik \& Ryde 1997; Esin, McClintock \&
Narayan 1997). Variations in the inner disc radius provide the
necessary change in the soft photon input that can explain the observed
rms($E$). However, Z02 analysed the long-term variability of Cyg X-1 on
timescales of days and months, much longer than the viscous timescale,
so significant variations in the inner disc radius are possible. This
is not the case on the timescales of milliseconds and seconds, observed
here. The radial drift velocity in the accretion disc at 0.05 of the
Eddington luminosity truncated at $20 GM/c^2$ around a $10 {\rm
M}_\odot$ black hole is $\sim$4 km s$^{-1}$ in the radiation pressure
dominated zone (Shakura \& Sunyaev 1973). The timescale required to
change the truncation radius by e.g.\ $10 GM/c^2$ is $\sim$20 s. When
the disc is truncated farther away, this timescale is even longer.
Clearly, there must be another mechanism for varying soft photon input.

Most of the rapid X-ray variability models assume some kind of
oscillations in the accretion flow. It is not yet clear how these
oscillations can be converted into the observed modulation of X-ray
flux. Our result suggests that in the hard state this happens via
modulation of the seed photons for Comptonization. In the model of
Giannios \& Spruit (2004), oscillations in the hot inner flow are
excited by variations in the Compton cooling rate. During oscillations,
the inner flow changes its Compton $y$ parameter, which results in the
pivoting of the Comptonized spectrum. This is in agreement with
observations and our rms($E$) models. The drifting-blob model of
B{\"o}ttcher \& Liang (1999), where the local seed photon input varies
as the blob travels though inhomogeneous hot flow, predicts increase of
the rms with energy, contrary to what is observed.

According to the truncated disc model in the soft (and probably very
high) state, the cold disc extends down to the marginally stable orbit
and the Comptonized emission originates from the active regions or
corona above the disc (e.g.\ Gierli{\'n}ski et al.\ 1999; Poutanen \&
Fabian 1999). The rms($E$) patterns quickly increasing with energy are
consistent with the stable disc and variable corona (see also Churazov,
Gilfanov \& Revnivtsev 2001). Coronal flares produce most of the power
at higher energies (B{\"o}ttcher, Jackson \& Liang 2003), so naturally
we expect most variability at photon energies $\ga$10 keV.

The two rms($E$) models we considered in Section \ref{sec:soft_state}
can be explained within the disc-corona geometry. The model with
varying Comptonized normalization (but not spectral shape) can
correspond to a scenario in which the covering fraction of the (patchy)
corona varies, due to e.g.\ new flares being formed, and the \lhs\ ratio for
each flare is roughly the same, so the spectral shape of Comptonization
does not vary. However, with changing number of flares (covering
fraction), the luminosity of Comptonization would change, hence the
observed high-energy variability. Since the uncovered fraction of the
disc changes as well, we should expect some variability in the disc.
The very small disc variability found in this model
(Fig.~\ref{fig:soft_rmspec}a) requires a small covering fraction of the
corona, $\la$10 per cent.

The model with varying $\ell_{\rm h}$ may correspond to changing power in the
corona without changing the covering fraction. In the soft state of
XTE~J1650--500, it required $r(\ell_{\rm h}) \approx 17\%$ and additional 7\%
variability in the disc (Fig.~\ref{fig:soft_rmspec}a). A quick estimate
shows that this is the level of variability is expected from
reprocessing of hard Comptonized photons in the disc. In the energy
spectrum shown in Fig.~\ref{fig:soft_rmspec}(b), the Comptonized
component luminosity, $L_{\rm h}$, is roughly 0.4 of the disc
luminosity, $L_{\rm s}$. If we consider a corona or active regions
above the disc, then we would expect less then a half of $L_{\rm h}$ to
be absorbed and re-emitted by the disc, which constitutes $\la$0.2 of
$L_{\rm s}$. Since the coronal variability is about 20\%, the expected
variability of the disc from reprocessing is $\la$4\%, in a rough
agreement with the observed rms of 7\%. The remaining fraction of
variability might be intrinsic to the disc.

\section{Reflection}
\label{sec:reflection}

\begin{figure}
\begin{center}
\leavevmode \epsfxsize=6cm \epsfbox{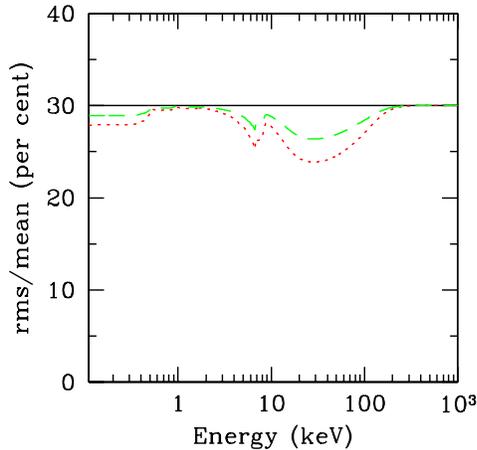}
\end{center}
\caption{The effect of Compton reflection on rms($E$) variability. The
energy spectrum used for this simulation is the best-fitting
model to the hard state. The curves represent a simple two-component
model, where the continuum and reflection vary in normalization only but
not in spectral shape. Solid curve: continuum and reflection are
correlated, and vary with the same relative amplitude of 30\%. Dotted
curve: both components are uncorrelated, and vary with the same
amplitude. Dashed curve: both components are correlated, but the reflection
variability amplitude is 15\%, i.e., a half of the continuum amplitude.}
\label{fig:ref}
\end{figure}

We have neglected the effects of Compton reflection in our variability
models so far. Here, we perform a simple test of possible effects
reflection variability can have on the rms spectra. To do this, we take
our best-fitting model to the hard state from Section
\ref{sec:hard_state} as a template. This time, we take into account
Compton reflection. We also assume the simplest possible two-component
variability model, where the continuum (with the soft and hard components added)
and the reflection vary only in normalization, but not in spectral shape.
When the continuum and reflection variabilities are correlated (i.e.\ the
covariance $\sigma_{\rm cr} = \sigma_{\rm c}\sigma_{\rm r}$) and vary
with the same relative amplitude, the resulting rms($E$) is obviously
energy-independent (solid curve in Fig.~\ref{fig:ref}). However, strong
reflection-related features appear in the rms spectrum (dotted curve in
Fig.~\ref{fig:ref}), when continuum and reflection are uncorrelated
($\sigma_{\rm cr} =0$). Similar features can be seen when they are
correlated, but vary with different relative amplitudes (30\% and 15\%
for the continuum and reflection, respectively; dashed curve in
Fig.~\ref{fig:ref}). Certainly, this is a very simplified model, but
similar features in rms($E$) are generated in our parameter-variability
models in all spectral states when we allow for the reflection
variability to be detached from the continuum variability, i.e.\
either uncorrelated, of different amplitude, or entirely independent.

We do not see any obvious such features in the observed rms spectra
(Fig.~\ref{fig:trans}) in any of the spectral states. This provides
with an important constraint on the relation between the irradiating
X-ray continuum and Compton reflection. As they are well correlated,
the reflected component must originate from Compton reprocessing of the
{\em observed\/} continuum. This also means that the reflection
variability (at the frequencies dominating the power spectra
used to the analysis, see Fig.\ \ref{fig:pds}) is due to variability in
the irradiating X-ray continuum and not due to changes in the
properties of the reflector (e.g.\ waves or warps in the disc).

We have pointed out in Section \ref{sec:introduction} that the rms($E$)
spectra presented here have been integrated over a rather wide
frequency range, (1/512)--128 Hz, whereas spectral dependence on
frequency is possible. It has been known that the amplitude of Compton
reflection can strongly decrease at high frequencies, $\ga$10 Hz, e.g.,
in the hard state of Cyg X-1 (Revnivtsev et al.\ 1999b; Gilfanov,
Churazov \& Revnivtsev 1999). However, those frequencies contribute
relatively little to the integrated PDS used by us (Fig.\
\ref{fig:pds}). On the other hand, we note that reflection amplitude
changing with frequency would still produce featureless rms($E$)
spectrum as long as the {\em fractional\/} variability of the
reflection remains roughly constant with changing frequency.

\section{Conclusions}
\label{sec:conclusions}

We have explained various patterns of the observed energy-dependent
X-ray variability in black hole binaries with a model in which the energy
spectrum varies in response to a changing physical parameter. Our
spectral model consisted of the disc emission and its hybrid
(thermal/non-thermal) Comptonization. In the hard state, we found the
decreasing rms($E$) consistent with variations in the seed photon input
(together with some variation in $\ell_{\rm h}$). In the soft and very
high states, the data were consistent with varying power released in the
hot Comptonizing plasma. Another model of the soft state involved
varying coronal luminosity, without changing spectral shape.

Our models predict a few important features of the rms($E$) spectra.
The break in the rms($E$) observed in the soft (and perhaps hard) state is
directly related to the seed photon temperature. We estimate it to
occur at $\sim 15 kT_{\rm s}$. Thus, rms spectra extending to lower
energies than those used in this paper might yield an important constraint on
the origin of the soft excess observed in the hard state. In the very
high state (and perhaps sometimes in the soft state), our models
predict a strong peak in rms($E$) at $\sim$30 keV, related to the
temperature of the thermal electrons in the hybrid plasma. If this peak
is confirmed with high-energy data, it would strongly support the
presence of hybrid electrons in the hot plasma. We stress that the
presence of the peak requires {\em hybrid}, not just power-law
electrons.

Lack of clear reflection features in the rms spectra implies that the
reflection and the X-ray continuum are well correlated and vary with
the same amplitude. Therefore, the reflected component originates from
the reflection of the observed continuum indeed, and its rapid
variability is due to changes in the irradiating continuum, and not due
to changes in the reflector properties. We stress, however, that this
result applies to the range of frequencies dominating the PCA power
spectra of the studied objects, i.e., $\la$10 Hz.

Z02 studied energy-dependent patterns of variability from Cyg X-1,
based on {\it RXTE}/ASM and {\it CGRO}/BATSE light curves on timescales
from days to months. In this work, we have extended this study to other
black hole candidates and to much shorter timescales, from milliseconds
to hundreds of seconds. The observed dependence of fractional rms
variability on energy is very similar in both cases, despite very
different timescales. Moreover, similar patterns have been observed
from PDS components, like QPOs (e.g.\ Rodriguez et al.\ 2004a, b;
Zdziarski et al.\ 2005), also in the case of neutron-star binaries
(e.g.\ Gilfanov et al.\ 2003). Clearly, there is a common physical
mechanism behind those rms($E$) patterns. Their universality indicates
their fundamental nature and importance for understanding physics of
accretion.

On the other hand, significant difference in timescales {\em implies\/}
different physics. While Z02 explained hard-state variability by
changing inner disc radius, it cannot operate on timescales of
milliseconds, which are much shorter than the viscous timescale. In
both cases the common underlying mechanism is modulation of hard X-rays
by the varying seed photon input, however the dynamical link between
the cold disc and hot Comptonizing region must be different.

\section*{Acknowledgements}

We thank Chris Done for stimulating discussions and the referee, Juri
Poutanen, for valuable suggestions. This paper was supported through
KBN grants 1P03D01827, 1P03D01727, PBZ-KBN-054/P03/2001 and 4T12E04727.
MG acknowledges support through a PPARC PDRF.


\label{lastpage}

\end{document}